\pdfoutput=1

\documentclass[10pt,a4paper]{article}

\usepackage[english]{babel}
\usepackage{amscd,amssymb}
\usepackage[fleqn]{amsmath}
\usepackage{bm}
\usepackage{lmodern,microtype}
\usepackage{hyperref}

\usepackage[left=4cm]{geometry}

 \renewcommand{\i}{\text{i}}

\newcommand{\be}{\begin{equation}}
\newcommand{\ee}{\end{equation}}
\newcommand{\ba}{\begin{eqnarray}}
\newcommand{\ea}{\end{eqnarray}}

\newtheorem{theorem}{Theorem}[section]

\newtheorem{conjecture}[theorem]{Conjecture}

\title{A staggered fermion chain with supersymmetry on open intervals}
\author{\textsc{Matteo Beccaria}\footnote{Dipartimento di Matematica e 
Fisica ``Ennio De Giorgi'', Universit\`a del Salento, Via Arnesano, 73100 Lecce \& INFN, Sezione di Lecce, Italy. Email: \href{mailto:matteo.beccaria@le.infn.it}{matteo.beccaria@le.infn.it}}, and \textsc{Christian Hagendorf}\footnote{ Section de Math\'ematiques, Universit\'e de Gen\`eve, 2-4, rue du Li\`evre, 1211
Gen\`eve 4, Switzerland. Email: \href{mailto:christian.hagendorf@unige.ch}{christian.hagendorf@unige.ch}}}
 \date{}
 
\begin{document}

\maketitle

\begin{abstract}
A strongly-interacting fermion chain with supersymmetry on the lattice and open boundary
conditions is analysed. The local coupling constants of the model are staggered, and the properties
of the ground states as a function of the staggering parameter are examined. In particular, a connection between certain ground-state components and solutions of non-linear recursion relations associated with the Painlev\'e VI equation is conjectured. Moreover, various local occupation probabilities in the ground state have the so-called scale-free property, and allow for an exact resummation in the limit of infinite system size.
\end{abstract}

\section{Introduction}

The understanding of strongly-interacting many-particle systems is one of the prime challenges of statistical mechanics and condensed matter physics. As realistic models are a posteriori notoriously difficult to handle analytically, one resorts often to simplified and mathematically more accessible descriptions which preserve essential physical features. The analytical accessibility is usually due to some underlying symmetry. In this work we study \textit{supersymmetry}, applied to models of strongly-repulsive itinerant fermions.

More precisely, we consider the class of $\mathcal M_\ell$ models which were introduced in \cite{fendley:03_2,fendley:03}. They describe spinless fermions on arbitrary lattices with the exclusion rule that connected particle clusters contain at most $\ell$ particles. The models have a built-in supersymmetry: the corresponding generators -- the supercharges -- add or remove single particles from the system while preserving the exclusion constraints. We focus on $\ell=1$ which corresponds to fermions with nearest-neighbour exclusion. To date, this is certainly the best studied case, and presents all features of a realistic physical model. 
The analysis of the one-dimensional $\mathcal M_1$-chain with periodic and open boundary conditions revealed interesting connections with enumerative combinatorics: in suitable normalisation, the ground state components are given by integers which enumerate alternating sign matrix number with various symmetries \cite{fendley:03, beccaria:05}, as are known from the properties of the XXZ spin chain at $\Delta=-1/2$ (see for example \cite{razumov:00,razumov:01,degier:02, difrancesco:06}). Indeed, a mapping between the models can be established, what led to discovery of a hidden supersymmetry in the spin chain \cite{yang:04}. In two dimensions, the model presents remarkable features  \cite{fendley:05,huijse:08_2} such as the so-called 'superfrustration', a phase with extensive ground-state entropy \cite{fendley:05_2}, and direct relations to rhombus tilings \cite{jonsson:06,jonsson:10, huijse:10_1}. For a comprehensive introduction we refer to \cite{huijse:10}.

The continuum limit of the $\mathcal M_1$ fermion chain is described by a superconformal field theory with central charge $c=1$ (a free boson with a special compactification radius) \cite{huijse:11_2}. In this sense, it is quantum critical. It was understood in \cite{fendley:10,fendley:10_1} that an off-critical extension can be achieved through the introduction of a staggering parameter on the lattice. The staggering is $3$-periodic, and closely related to the so-called $3$-rule appearing in cohomology considerations about the model's ground states, and therefore these works focused on periodic lattices with the number of sites being a multiple of three. The ground states are polynomials in the staggering parameter. In particular, it was pointed out that several of its polynomial components appear also in the exact ground states of the XYZ spin chain along a special line of couplings, and are solutions to a non-linear recursion relations related to the Painlev\'e VI differential equation \cite{bazhanov:06,mangazeev:10,zinnjustin:12}. Further evidence for the connection between the staggered $\mathcal M_1$ model and the special XYZ chain was given \cite{hagendorf:12}: in fact, their Hamiltonians have coinciding spectra in several momentum subsectors.

In this paper, we consider the staggered $\mathcal M_1$ model on a one-dimensional lattice with open boundary conditions. This problem has already been addressed in \cite{huijse:11}, where several properties of the ground states were investigated through perturbation theory around trivially solvable points. In particular, the presence of kinks separating two types of ordered states was pointed out. The main purpose of the present article to expand upon these results, and show that the finite-size ground states have a number of remarkable non-perturbative properties. In fact, like in the case of periodic systems we conjecture a relation to polynomials pertaining to hierarchies of integrable equations, this time however in a more refined form. In particular, we point out a variety of sum rules for the square norm of the ground state vectors. Moreover, we exploit the so-called \textit{scale-free property} of local observables discovered in \cite{fendley:10, fendley:10_1}: it states that their Taylor/perturbation expansion in the staggering parameter around trivial points can be summed, and thus yield a conjectured exact and non-perturbative form for infinite systems.

This article is organised as follows. In section \ref{sec:model} we introduce
the details of the $\mathcal M_1$ model, and describe different interesting variants for the staggering and number of sites. These
variants are investigated in sections \ref{sec:3n} and \ref{sec:3n1}: we
conjecture a relation between certain ground-state components of the models and solutions of the Painlev\'e VI-Hirota equations, and
provide evidence for the scale-free property of various quantities. We
present our conclusions in section \ref {sec:concl}.

\section{The model and methods}
\label{sec:model}
In this section we recall the general definition of the staggered $\mathcal M_1$ model for fermions with hard-core exclusion. Moreover, we explain an iterative approach to the determination of its ground states, exploiting the fact that it is a polynomial in the staggering parameter. The key polynomials appear to be deeply related to the theory of the Painlev\'e VI equation, and were first discovered in the context of a particular XYZ chain \cite{bazhanov:06,mangazeev:10}. We recall briefly their main features.

\subsection{Definition of the staggered $\mathcal M_1$ model}
We consider a one-dimensional chain with $N$ lattice sites which we label by integers $j=1,2,\dots, N$. The model describes spinless fermions living on this chain. These are created and annihilated by operators $c_j,c_j^\dagger $ which obey the usual anti-commutation rules $\{c_i,c_j\} = \{c_i^\dagger,c_j^\dagger\}=0$ and $\{c_i,c_j^\dagger\}=\delta_{ij}$.
The particle configurations are constrained by a nearest-neighbour
exclusion: at most one of two adjacent sites may be occupied. We will represent an
occupied site by $1$, and an empty site by $0$. For example, a configuration for
a six-site chain with particles on sites $2$ and $6$ is given by $\alpha=010001$.
The hard-core exclusion forbids therefore pairs like $\cdots 11 \cdots$. We will
sometimes abbreviate the empty configuration by $\bm 0 =00\cdots 0$. We shall work
with free boundary conditions what is equivalent to add two inaccessible but empty
sites at $j=0$ and $j=N+1$.


Our aim is to study a Hamiltonian generated by the two
supercharges
\begin{equation*}
Q = \sum_{j=1}^N \lambda_j d_j,\quad Q^\dagger =
\sum_{j=1}^N \lambda_j^\ast d_j^ \dagger,
\end{equation*}
where $d_j= P_{j-1}c_j
P_{j+1}$, with the projector $P_j = 1-c^\dagger_j c_j$, are dressed fermion
operators which respect the hard-core exclusion. As the sites $0$ and $N+1$ are
always empty, we set $P_0=P_{N+1}\equiv 1$. The $\lambda_j,\,j=1,\dots,N$ are
arbitrary complex numbers which we call sometimes coupling constants. The
supercharges are nilpotent for any choice of the $\lambda_j$'s:
$Q^2=(Q^\dagger)^2=0$. The Hamiltonian is given by $H=\{Q,Q^ \dagger\}$. In terms
of the fermions the $H$ it can be written as
\begin{equation*}
H = \sum_{j=1}^{N-1}
P_{j-1}\left(\lambda_{j+1}\lambda_j^\ast c_{j+1}^\dagger c_j+
\text{h.c.}\right)P_{j+2}+ \sum_{j=1}^N|\lambda_j|^2P_{j-1}P_{j+1}.
\end{equation*}
$H$ commutes with the fermion number operator $F = \sum_{j=1}^N c_j^\dagger c_j$,
and of course the supercharges. The algebraic relations of the quadruple
$H,Q,Q^\dagger,F$ yields the well-known $\mathcal N =2$ supersymmetry algebra \cite{witten:82}.

In this article we concentrate on the case where the number of sites is of the form
$N=3n$ or $N=3n-1$. For both choices and non-zero coupling constants the model has a single zero-energy ground state
in the subspace of $n$ particles as can be shown by cohomology arguments \cite{huijse:10_1}. 
Conversely, for $N=3n-2$ no such zero-energy ground state exists: supersymmetry is spontaneously 
broken. The proof of this statement uses a one-to-one correspondence between linearly independent ground 
states and the elements of the quotient space (or cohomology) $\mathfrak H_Q=\text{ker } Q/\text{im } Q$. Its dimension 
coincides with the dimension of the ground state space. In order to determine $\mathfrak H_Q$, one 
proceeds into two steps. The lattice is divided into two disjoint parts, call them $S_1$ and 
$S_2$. As the supercharges are linear superpositions of fermion annihilators/creators, we divide $Q=Q_1 + Q_2$ so that $Q_{1/2}$ acts only on $S_{1/2}$. First, one evaluates $\mathfrak H_{Q_1}$. Second one
acts with $Q_2$ within this space in order to find $\mathfrak H_{Q}$, applying the so-called
\textit{tic-tac-toe lemma}. Even though the details of this procedure are not important for our considerations, we 
wish to point out that the division into sublattices uses a fundamental feature of the 
$\mathcal M_1$ model, namely an underlying periodicity with period $3$. Indeed, in the existence proof $S_1=
\{2,5,8,\dots\}$ is chosen to contain every \textit{third} site, and $S_2$ all the remaining sites.  
This inherent ``3-rule'' led to the idea to choose the coupling constants $\lambda_j$ of the 
supercharges from a $3$-periodic pattern \cite{fendley:10,fendley:10_1}, namely $\lambda_1=y,\,
\lambda_2=1,\,\lambda_3=y,\dots$, and investigate their properties as a function of the real 
parameter $y$. For periodic chains with $N=3n$ sites, it does not matter which one of the three 
coupling constants is scaled to one. For open chains however, various choices lead to different situations. We choose to investigate two one-parameter staggerings:
\begin{align}
  \text{(I)}\quad \lambda_{3p-2}=y,\,\lambda_{3p-1}=1,\,\lambda_{3p}=y, \label
{eqn:stag1}\\ 
\text{(II)}\quad \lambda_{3p-2}=y,\,\lambda_{3p-1}=y,\,\lambda_{3p}=1,
\label {eqn:stag2}
\end{align}
where $p=1,\dots,n$. Notice that the choice (I) is invariant
under a parity operation $P$ which reverses the order of all sites\footnote{We leave aside the 
question of possible signs through the reordering of fermions.}. For this
choice, the Hamiltonian commutes with the parity operator $[H,P]=0$, and eigenstates may be
classified as parity-even and parity-odd. Under parity the staggering pattern
$(y,y,1)$ of (II) is mapped to (II') $(1,y,y)$, whose properties can readily be
deduced from (II), and will therefore not be considered separately.
As $y=1$ all the cases reduce to the quantum critical model studied in \cite{beccaria:05}. Here, we focus on general values for $y$.

\subsection{Polynomiality and diagonalisation methods}

Our aim is to find the properties of the zero-energy ground states.
It is natural to choose to do this in the occupation number basis. Therefore we expand
the single solution to solving $H(y)|\Psi(y)\rangle = 0$ according to
\begin{equation*}
  |\Psi(y)\rangle =\sum_\alpha \psi_\alpha(y)|\alpha\rangle.
\end{equation*}
Here $\alpha$ runs over all admissible configurations on $N$
sites with $n$ particles, where $N=3n$ or $N=3n-1$ according to the case under consideration. For both choices (I) and (II) of the staggering, the
supercharges are linear in the coupling constant $Q=Q_0+yQ_1$, and therefore the
Hamiltonian is a quadratic polynomial in $y$: $H(y) = H_0+y H_1 + y^2 H_2$. This
implies that we can choose the normalisation for $|\Psi(y)\rangle$ in such a way
that all components $\psi_\alpha(y)$ are polynomials. As this choice might still
contain some redundancies, we impose that for any given configuration $\alpha$
there is at least one $\alpha'$ such that the polynomials $\psi_\alpha(y)$ and
$\psi_{\alpha'}(y)$ have no common factors (i.e they are coprime). This fixes the
ground states up to an overall $c$-number normalisation which will be adjusted
accordingly.

The actual construction of  $|\Psi(y)\rangle$ is based on the following observation. 
Since all $\psi_{\alpha}(y)$ are polynomials, we can expand
\begin{equation*}
|\Psi(y)\rangle = \sum_{\ell=0}^{d} y^{\ell}\,|\Psi_{\ell}\rangle,
\end{equation*}
for a certain degree $d$, depending on the lattice size $N$.
In all the cases studied below, we provide the explicit form of the zero-order
term $|\Psi_{0}\rangle$ which is obtained in the limit $y\to 0$. Of course, this is a well defined state, 
although the zero energy space can be degenerate at $y=0$. The above decomposition can also be written
as 
\begin{equation*}
|\Psi(y)\rangle = m(y)\,|\Psi_{0}\rangle+|\widetilde\Psi(y)\rangle,
\end{equation*}
where $m(y) = 1+\sum_{\ell=1}^{d'}m_{\ell}\,y^{\ell}$, for a certain degree $d'$,  and $\langle \Psi_{0}|\widetilde\Psi\rangle = 0$. In all cases, we provide a conjecture for the polynomial $m(y)$ (again, depending on the lattice size). This means that the identification of the non-degenerate state obeying $H\,|\,\Psi(y)\rangle=0$, for generic $y$, will reduce to the solution of
\begin{equation*}
H_{0}\,|\Psi_{\ell}\rangle+
H_{1}\,|\Psi_{\ell-1}\rangle+
H_{2}\,|\Psi_{\ell-2}\rangle=0,
\end{equation*}
where the ambiguity in the component of $\ker  H_{0}$ is solved in terms of $m(y)$. Indeed, by consistency we have 
\begin{equation*}
H_{1}\,|\Psi_{\ell-1}\rangle+
H_{2}\,|\Psi_{\ell-2}\rangle \perp \ker H_{0},
\end{equation*}
and we can write 
\begin{equation*}
|\Psi_{\ell}\rangle = m_{\ell}\,|\Psi_{0}\rangle-H_{0}^{-1}\left[
H_{1}\,|\Psi_{\ell-1}\rangle+
H_{2}\,|\Psi_{\ell-2}\rangle\right].
\end{equation*}
This simple algorithm can be applied starting with $|\Psi_{-1}\rangle = |\Psi_{-2}\rangle = 0$ and, by 
consistency, terminates as soon as $|\Psi_{\ell}\rangle = |\Psi_{\ell-1}\rangle = 0$ for a certain $\ell=d$. Here $H_0^{-1}$ is the inverse on the orthogonal complement of $\ker H_0$. Its computation is trivial, since $H_{0}$ is diagonal in the occupation number basis. Moreover, we exploit sparse-matrix linear algebra in order to apply $H_{1}$ and $H_{2}$.

\subsection{Relation to the XYZ chain and Hirota equations}

It was pointed out in \cite{fendley:10,fendley:10_1} that the ground states of the staggered fermion chain with periodic boundary conditions and $N=3n$ sites present some striking similarities to the ground states of the XYZ Hamiltonian
\begin{equation*}
	H_{\text{XYZ}}=-\frac{1}{2}\sum_{k=1}^L \sum_{a=x,y,z} J_a \sigma_k^a\sigma_{k+1}^a.
\end{equation*}
Here $\sigma^a,\,a=x,y,z$ denote the usual Pauli matrices. We choose periodic boundary conditions, and 
restrict the coupling constants $J_a$ to the
the line
\begin{equation*}
	J_x=1-\zeta,\quad J_y=1+\zeta, \quad J_z=\frac{1}{2}(\zeta^2-1).
\end{equation*}
For $L=2n+1$ sites this chain has exactly two translationally-invariant ground states $|\Phi_\pm\rangle$ at the energy $E_0=-L(\zeta^2+3)/4$ \cite{hagendorf:12_1}. These ground states are related by the reversal of all spins, and therefore it is sufficient to study one of them, say $|\Phi_-\rangle$ which contains the configuration $|{\downarrow\cdots \downarrow}\rangle$ but has zero overlap with $|{\uparrow\cdots \uparrow}\rangle$. Bazhanov and Mangazeev showed that under suitable normalisation the component in front of the completely polarised configuration $\downarrow\downarrow\cdots\downarrow$ can be written as
\begin{equation*}
  \phi_{\downarrow\downarrow\cdots\downarrow}(\zeta)=\zeta^{n(n+1)}s_n(\zeta^{-2})
\end{equation*}
where the $s_n(z), \, n\in \mathbb C$ are polynomials in $z$ of degree $\lfloor n^2/4\rfloor$ which solve the non-linear differential recursion relation
\begin{align} &2z(z-1)(9z-1)^2(\ln s_n(z))'' + 2(3z-1)^2(9z-1)(\ln
s_n(z))'\nonumber\\ &+8(2n+1)^2 \frac{s_{n+1}(z)s_{n-1}(z)}{s_n(z)^2}
-(4(3n+1)(3n+2)+(9z-1)n(5n+3))\nonumber \\
&=0, \label{eqn:hirota}
\end{align} with initial
conditions $s_0(w)=s_1(w)\equiv 1$. This recursion is a special case of the Hirota equations satisfied by a tau-function hierarchy associated to the Painlev\'e VI differential equation \cite{okamoto:87, bazhanov:06}.

These polynomials appear also in the two zero-energy ground states of the staggered fermion chain at $N=3n$ sites with periodic boundary conditions. The precise relation is as follows. The ground states can be chosen to be parity eigenstates. Write them as $|\hat \Psi^\pm(y)\rangle$ where the label $\pm$ corresponds to the parity eigenvalue. Upon appropriate normalisation we find that the components of the configuration $010010\cdots 010$ are related to the $s$-polynomials according to
\begin{align*}
	\hat\psi_{010\cdots 010}^+(y) &=\frac{\zeta^{\lfloor(n-1)^2/2\rfloor }}{2^{(n-1)(n-2)/2}}\frac{s_{n-1}(\zeta^{-2})}{s_{n-1}(0)},\\
	\hat\psi_{010\cdots 010}^-(y) &=\frac{\zeta^{\lfloor n^2/2\rfloor }}{2^{(n-1)(n+2)/2}}\frac{s_{-n}(\zeta^{-2})}{s_{-n}(0)}.
\end{align*}
where the variables $y$ and $\zeta$ are related through the quadratic equation
\begin{equation*}
  \zeta^2=1+8y^2.
\end{equation*}
This hints strongly at a profound relation between the two models, but a precise mapping is so far only known in the critical case $y=1$, $\zeta=3$.

\subsection{Factorisation of Hirota polynomials}
In this work, we will point out a connection of the open staggered chains to the Hirota polynomials. It is more subtle than for periodic boundary conditions, and related to factorisation properties of the $s_n(z)$. These were conjectured in \cite{mangazeev:10}, and here we quote this conjecture:

\begin{conjecture}

For all $n\in\mathbb{Z}$ the ratios $s_n(z)/s_{n}(0)$
are polynomials with integer coefficients. Also, for all $n\ge -1$, $s_{n}(0) = 1$.
For any $n$ they factorise into two
polynomials according to the following scheme:

\begin{enumerate}
	\item For $n=2k+1, k\in\mathbb{Z}$, we have 
	\begin{equation}
		s_{2k+1}(w^2) = s_{2k+1}(0)p_k(w)p_k(-w),\quad p_k(0)=1.
		\label{eqn:defp}
    \end{equation} 
    where the $p_k(w)$ are
	polynomials with $\deg p_k(w)=k(k+1)$ and integer coefficients. We have the
	transformation property
	\begin{equation*}
		p_k(w)=\left(\frac{1+3w}{2}\right)^{k(k+1)}p_k\left(\frac{1-w}{1+3w}\right).
	\end{equation*}

	\item For $n=2k, k\in\mathbb{Z}$, the polynomial factorises according to
	\begin{equation}
		s_{2k}(w^2) = c_k(1+3w)^{k(k+1)}p_{-k-1}\left(\frac{w-1}{1+3w}\right)q_{k-1}(w),\quad q_k(0)=1.
		\label{eqn:defq}
	\end{equation}
	with $c_k=2^{-k(k+2)}$ if $k\geq 0$, and $2^{-k^2}(2/3)^{2k+1}$ otherwise. 
	The polynomials $q_k(w)$
have $\deg q_k(w)=k(k+1)$ and positive integer coefficients. Moreover, they are
even and obey the transformation rule
\begin{equation*}
	q_k(w)= \left(\frac{1+3w}{2}\right)^{k(k+1)}q_k\left(\frac{1-w}{1+3w}\right).
\end{equation*}
\end{enumerate}
\end{conjecture}

A list of the first polynomials is given in the appendix.
In the next sections, we will present several conjectures about the ground states of the open staggered fermion chain. In particular, we claim that the polynomials $p_k(w)$ and $q_k(w)$ appear in various disguises, not only as components in the ground state vectors but also in their square norms, and projections on the ground states at weak coupling $y=0$ and strong coupling $y\to \infty$.

\section{Ground-state properties of the staggered chain with ${N=3n}$ sites}
\label{sec:3n}

We start with staggered open chain for
the two choices (I) and (II) for $N=3n$ sites. In both cases, we first determine
systematically the ground states at $y=0$ and $y\to \infty$ (weak and strong
coupling limits). Next, we present results at finite coupling $y$ from exact
diagonalisation of small systems which we use as a basis for our conjectures. Finally, we study site occupation probabilities near the boundaries and in the bulk, point out scale-free expansions in around the asymptotic points, and use them to conjecture exact expressions in the limit of infinite systems.

\subsection{Staggering I}
\subsubsection{Asymptotic ground states}
\label{sec:asymy1y}
We begin with the derivation of the $y=0$ ground state, which turns out to be a kink state as was shown in \cite{huijse:11} by means of perturbation theory. Here we prefer rather to exploit the properties of the supercharges $Q,Q^\dagger$ alone. The considerations are very reminiscent of the $3$-rule, and use polynomiality/analyticity in $y$. Moreover, they will prove to work also for staggering II (see section \ref{sec:3n1yy1}).

At $y=0$, we have the supercharge $Q_0=\sum_{j=1}^n d_{3j-1}$. As it acts on isolated sites, it
is clear that any state annihilated by this operator must not contain any particles
on the sites $3j-1$ with $j=1,\dots,n$. For $Q_0^\dagger$ to annihilate such a state however, there must a least
be one particle adjacent to such a site. This requirement is fullfilled by all states of the
form
\begin{equation*} 
	|m\rangle=|\underset{3m}{\underbrace{100100\cdots
100}}\underset{3(n-m)}{\underbrace {001001\cdots 001}}\rangle, \quad m=0,\dots,n
\end{equation*} 
These are \textit{kink states} in the sense that they interpolate
between two ordered configurations $100100\cdots$ and $\cdots 001001$. All of them
are admissible ground states at $y=0$, and therefore $E=0$ has an $(n+1)$-fold
degeneracy. The general solution is a superposition $|\Psi(y=0)\rangle =
|\Psi_K\rangle= \sum_{m=0}^na_m|m \rangle$. In order to determine the relevant
combination we demand \textit{continuity} with $y \neq 0$, where we know that only a single
ground state may exist. The polynomial character of the ground-state wave function
allows to write $|\Psi(y)\rangle=|\Psi_K \rangle+y|\Psi_1\rangle + \dots$ Acting
with the supercharges leads to the equations $Q_0|\Psi_1\rangle +
Q_1|\Psi_K\rangle=0$ and $Q_0^\dagger|\Psi_1\rangle + Q_1^\dagger| \Psi_K\rangle=0$
at first order in $y$. We claim that the second equation is already sufficient to
determine $|\Psi_K\rangle$, even without knowing the structure of $|
\Psi_1\rangle$. To show this, we examine the action of $Q_1^\dagger$ on the states
$|m \rangle$. Let us introduce the defect states
\begin{equation*}
	|\phi_m\rangle=|\underset{3m}{\underbrace{100\cdots 100}}101\underset{3(n-m-1)}
{\underbrace{001\cdots 001}}\rangle,\quad m=0,\dots ,n-1
\end{equation*}
with $n+1$ particles. Then we have
\begin{equation*}
	Q_1^\dagger |m\rangle = (-1)^m\times\left\{
	\begin{array}{ll}
		|\phi_0\rangle, & m=0,\\
		|\phi_{n-1}\rangle & m=n,\\
		|\phi_{m-1}\rangle+|\phi_m\rangle, & \text{otherwise}.
	\end{array}
	\right.
\end{equation*}
None of the states $|\phi_m\rangle$ contains any particles
on the sites $3j-1$, and therefore none of them can be obtained by acting with $Q_0^\dagger$ on any
other state having $n$ particles. So we must have $Q_0^\dagger|\Psi_1\rangle=0$.
This implies the additional requirement that $Q_1^\dagger |\Psi_K\rangle=0$. More
explicitly, we find
\begin{equation*}
	\sum_{m=0}^{n-1}(-1)^m(a_m-a_{m+1})|\phi_m\rangle =0.
\end{equation*}
As the $|\phi_m\rangle$ are linearly independent, we conclude that $a_0=a_1=a_2=
\dots=a_n$. This determines already the ground state up to an overall normalisation
because for $y>0$ it is unique. Setting the normalisation to unity we thus find the ground state
\begin{equation*}
  |\Psi_K\rangle = \sum_{m=0}^n|m\rangle,
\end{equation*}
which is a uniform superposition of kinks. This is the result of \cite{huijse:11}.

As $y\to \infty$, the ground state has to be annihilated by $Q_1$ and its adjoint
$Q_1^ \dagger$. Let us write this in an suggestive way: $Q_1 = d_1
+\sum_{j=1}^{n-1}(d_{3j}+d_ {3j+1})+d_{3n}$. From this it is clear that any state
annihilated by both these supercharges must not have a particle on the boundary
sites $1$ and $3n$, but a particle on their neighbours $2$ and $3n-1$ respectively.
Because of the hard-core exclusion, the situation obtained in this way is equivalent to
the initial one with six less sites (unless one starts with only three sites where the
problem is trivial), and one can proceed by iteration. The ground state for $y\to
\infty$ will therefore be proportional to the polarised state $010010\cdots 010$:
\begin{equation}
	|\Psi_P\rangle=\sum_{j=1}^n c_{3j-1}^\dagger|\bm{0}\rangle.
	\label{eqn:polarised}
\end{equation}

\subsubsection{Finite coupling}
Now we turn to finite values for the coupling constant $y$. We list some properties, based on observations from exact diagonalisation of the Hamiltonian
for small system up to $N=24$ sites ($n=8$ particles). In order to give a simple illustration, we display here the ground state components for six sites. As the Hamiltonian commutes with parity it is sufficient to indicate the representatives
\begin{align*}
	\begin{array}{llll} 
		101000 & -y^2 & 100100 & 1+3 y^2\\ 010100 & -y\left(1+4 y^2\right) & 100010 & -2 y \left(1+2 y^2\right)\\
010010 & y^2 \left(3+8y^2\right) \qquad & 100001 & 1+2 y^2
	\end{array}
\end{align*}
The ground state wave function is in fact even under the parity operation. It is not difficult to see that this holds for any $n$: the asymptotic ground states at $y\to 0$ and $y\to \infty$ are parity-even. Since the eigenvalues of $P$ are discrete this must also hold at finite coupling. From our example we observe that all components have definite transformation behaviour under $y\to -y$: if $S_\alpha$ denotes the total number of particles on the sites $2,5,8,\dots, 3n-1$ in configuration $\alpha$ then we have $\psi_\alpha(-y)=(-1)^{S_\alpha}\psi_\alpha(y)$. This is related to the fact that the unitary operator $U=\exp \i \pi \sum_{j=1}^n c_{3j-1}^\dagger c_{3j-1}$ has the property $UH(y)U^{-1} = -H(-y)$. Taking into account the uniqueness of the ground state and the boundary condition at $y=0$, we conclude $|\Psi(-y)\rangle = U|\Psi(y)\rangle$.

Let us now have a closer look at certain components. To motivative this, we compute for our example of six sites the square norm of the ground state vector, as well as its projections on the asymptotic ground states at $y=0$ and $y\to \infty$:
\begin{align*}
    &||\Psi(y)||^2=\left(3+8 y^2\right) \left(1+6 y^2+11 y^4+8
	y^6\right),\\
	&\langle \Psi_P|\Psi(y)\rangle= y^2(3+8y^2),\quad \langle \Psi_K|\Psi(y)\rangle=3+8y^2.
\end{align*}
The polynomial $3+8y^2$ appears thus in multiple locations. It is one of the two factors of the square norm. The other factor can in fact be written in terms of the polynomial $q_2(w)$ related to the Hirota equation, after a suitable change of variables to $\zeta=\sqrt{1+8y^2}$. The systematic examination of the ground states up to $n=8$ leads to the following

\begin{conjecture}
All ground state components are polynomials with integer coefficients and definite parity under $y\to -y$. The projections on the kink state and the polarised state differ only by a monomial
\begin{align*}
  \langle\Psi_P|\Psi(y)\rangle=y^n\langle \Psi_K|\Psi(y)\rangle
\end{align*}
Here $\langle \Psi_K|\Psi(y)\rangle$ is an even polynomial in $y$ of degree $n(n-1)$ with integer coefficients. The square norm factorises according to
\begin{equation*}
	  ||\Psi(y)||^2=\langle \Psi_K|\Psi(y)\rangle \times \left(\frac{\zeta}{2}\right)^{n(n-1)}q_n(\zeta^{-1}), \quad \zeta^2=1+8y^2,
\end{equation*}
where $q_n(w)$ is the polynomial implicitly defined in \eqref{eqn:defq}.
\end{conjecture}
The general form of the ground state is
\begin{equation*}
|\Psi(y)\rangle = \sum_{k=0}^{n} m_{k}(y) \, |k\rangle + |\widetilde \Psi\rangle, \qquad \langle k|\widetilde\Psi\rangle=0,
\end{equation*}
with certain polynomials $m_k(y)$, and $|k\rangle$ the individual kink states defined in section. We normalise them according to $m_{k}(y) = 1+\mathcal{O}(y^{2})$. They are fixed by the requirement that the 
iterative iterative algorithm for the construction of $|\Psi(y)\rangle $ terminates.

We have not found a relation between the projections on the kink states, and the polynomials coming from the Hirota equation \eqref{eqn:hirota}. Here we list $t_n(y)=\langle \Psi_K|\Psi(y)\rangle$ for $n=1,\dots,5$:
\begin{align*}
  t_1(y) =& \, 2,\\
  t_2(y) =& \, 3+8y^2,\\
  t_3(y) =& \, 2 \left(1+4 y^2\right) \left(2+7 y^2+8 y^4\right),\\
  t_4(y) =& \, 5+72 y^2+420 y^4+1300 y^6+2304 y^8+2304 y^{10}+1024 y^{12},\\
  t_5(y) =& \,2(3+70 y^2+720 y^4+4332 y^6+17020 y^8+46053 y^{10}+87840 y^{12}\\
  &+118080 y^{14}+108288 y^{16}+61440 y^{18}+16384 y^{20}),
\end{align*}

\subsubsection{Densities}
We probe the system by examination of the probability to find a particle at site $j$:
\begin{equation*}
  \rho_n^{(j)}(y) =\langle
\Psi(y)| c_j^\dagger c_j|\Psi(y)\rangle/||\Psi(y)||^2.
\end{equation*}
In particular, at the boundary $j=1$ we know the result for any $n$ at the critical point $y=1$ \cite{beccaria:05}:
\begin{equation*}
  \rho_n^{(1)}(y=1)=\frac{n(10n+11)}{2(4n+3)(4n+5)}
\end{equation*}
Therefore, in the limit of large systems $n\to \infty$ we find the density $5/16$.

For arbitrary $y$ the densities are given by ratios of polynomials at finite $n$. While we do not have closed-form expressions for these we examine their series expansions around the points $y=0$ and $y=\infty$, by following the strategy of \cite{fendley:10_1}. There it was pointed out that these series expansions have a so-called scale-free property: the first $\sim N$ coefficients do not depend on $N$. Let us examine this for the boundary density at site $j=1$ near the strong coupling point $y=\infty$. Up to $n=6$ the Taylor
expansions of $4y^2 \rho_n^{(1)}(y)$ are given by
\begin{equation*}
\begin{array}{l}
4y^2\rho_1(y)=\bm{1}-\frac{1}{2 y^2}+\frac{1}{4 y^4}-\frac{1}{8 y^6}+\frac{1}{16
y^8}+ \dots \\ 4y^2\rho_2(y)=\bm{1+\frac{1}{8 y^2}}-\frac{39}{64
y^4}+\frac{321}{512 y^6}-\frac{1735} {4096 y^8}+\frac{6209}{32768
y^{10}}-\frac{5671}{262144 y^{12}}+\dots\\ 4y^2\rho_3(y)=\bm{1+\frac{1}{8
y^2}+\frac{3}{64 y^4}}-\frac{417}{512 y^6}+\frac{5347} {4096
y^8}-\frac{39505}{32768y^{10}}+\frac{201147}{262144 y^{12}}+\dots\\
4y^2\rho_4(y)=\bm{1+\frac{1}{8 y^2}+\frac{3}{64 y^4}+\frac{3}{128 y^6}}-\frac{2403}
{2048 y^8}+\frac{42355}{16384y^{10}}+\frac{101479}{32768 y^{12}}+\dots\\
4y^2\rho_5(y)=\bm{1+\frac{1}{8 y^2}+\frac{3}{64 y^4}+\frac{3}{128 y^6}+\frac{7}{512
y^8}}-\frac{29249}{16384y^{10}}+\frac{659389}{131072 y^{12}}+\dots\\
4y^2\rho_6(y)=\bm{1+\frac{1}{8 y^2}+\frac{3}{64 y^4}+\frac{3}{128 y^6}+\frac{7}{512
y^8}-\frac{9}{1024y^{10}}}-\frac{185329}{65536 y^{12}}+\dots \end{array}
\end{equation*}
The bold-printed terms correspond to the stable part of the series, i.e. they are not changed as $n$ increases. The stable part of the series matches the series expansions
\begin{equation*}
\rho_\infty^{(1)}(y)=\frac{1+12 y^2-8 y^4+8y \left(y^2-1\right)^{3/2}}{16 y^2},\quad y\geq 1.
\end{equation*}
We conjecture that this is indeed the thermodynamic limit of $\rho_n^{(1)}(y)$, however the singularity at $y=1$ with critical exponent $3/2$ indicates that this can only hold for $y\geq 1$. This conjecture is in fact supported by the observation that this function reproduces the correct asymptotic value
$\rho_\infty(y=1)=5/16$ at the critical point.

Around $y=0$ we observe that the series expansions of boundary density does not have the scale-free property. We attribute this to the non-local nature of the kink state, as opposed to the very local and ordered character of the polarised state at $y\to \infty$. The first coefficients match the pattern
\begin{equation*}
  \rho_n^{(1)}(y)=\frac{n}{n+1}\left(1-\frac{n+3}{n+1}y^2+\dots\right)
\end{equation*}
Yet we are able to guess the limit as $n\to \infty$. To this end let us anticipate a little. In the next section we study the chain with staggering (I) and the number of sites $N=3n-1$. As it will turn out, that case has a simple ordered state at $y=0$, and allows to evaluate the corresponding boundary density $\rho_n^{(1)}(y)$ near $y=0$ from a scale-free expansion. In the limit of infinite system size $n\to \infty$ the expressions in these two cases should match as the other boundary is very far away. Thus we conjecture that also in the present case $N=3n$ the limit $n\to \infty$ yields
\begin{equation*}
  \rho_\infty^{(1)}(y)= \frac{1+44y^2-8y^4-(1+8y^2)^{3/2}}{32y^2},\quad y\leq 1.
\end{equation*}

Next, we discuss the centre of the chain. In the limit of large systems, we
expect the average values of local densities to measured in the bulk to be insensible to the boundary.
Therefore it is reasonable to speculate that some of them will be related to local densities
of the closed chain with periodic boundary conditions in the thermodynamic limit. We probe the density at the site $j_c=3n/2-1$ for even $n$, and $j_c=(3n+1)/2$ for odd $n$. Despite the presence of the kink state in the weak-coupling regime we find scale-free expansions. Extracting the pattern for the stable part, we conjecture the following expressions
  \begin{equation*}
  \rho^{(j_c)}_\infty(y) =
  \begin{cases}
      \frac{1+2y^2-\sqrt{1+8y^2}}{2(y^2-1)}, & y\leq 1\\
    \frac{4y^2-1+4y\sqrt{y^2-1}}{1+8y^2}, & y\geq 1
  \end{cases}.
\end{equation*}
These correspond to the staggered densities of the closed chain in the thermodynamic limit, for the parity-even ground state \cite{fendley:10_1}. The parity sector is expected because in the present case the open chain ground state is parity-even, too, as shown above.

\subsection{Staggering II}
\subsubsection{Asymptotic ground states}
\label{sec:asymyy1}
Let us first evaluate the ground states in
the limits $y=0$ and $y\to \infty$. At $y=0$ the ground state has a simple
structure as opposed to the kink state of case (I). We have $Q(y=0)=Q_0
=\sum_{j=1}^n d_{3j}$. For $Q_0$ to annihilate a state, there must not be any
particle on the sites $3j,\,j=1,\dots,n$. Conversely, if $Q_0^\dagger$ is to annihilate such a state,
each site $3j$ must have at least one neighbour occupied by a particle. All sites
$3j$ have two neighbours except for the last one $3n$. Hence there must be a particle
on site $3n-1$, and by iteration it is not difficult to see that this implies that all
sites $3j-1$ have to be occupied. Hence a state annihilated by both $Q_0$ and
$Q_0^\dagger$ is given by $010010010\cdots$:
\begin{equation*}
|\Psi_P\rangle =
|\Psi(y=0)\rangle = \prod_{j=1}^n c_{3j-1}^\dagger|\bm{0}\rangle.
\end{equation*}

Second, as $y\to
\infty$ the ground state must be annihilated by $Q_1=\sum_{j=1}^n
d_{3j-2}+d_{3j-1}$, and its Hermitian conjugate $Q_1^\dagger$. In the sector with
$N=3n$ sites and $n$ particles, $Q_1^\dagger$ annihilates any configuration that
contains a particle in each on the $n$ pairs of sites $(3j-2,3j-1),\,j=1,\dots,n$.
As these are disjoint, we can solve the problem for $Q_1$ separately for each such
pair. Consider thus the sites $1$ and $2$, with adjacent sites $0$ and $3$ being
empty. The only way to produce a state that is annihilated by $d_1+d_2$ with a
single particle is given by the singlet $\cdots 10\cdots-\cdots 01 \cdots$. We
conclude that the ground state becomes a tensor product of such singlets
\begin{equation*}
  |\Psi_S\rangle =
\prod_{j=1}^n(c^\dagger_{3j-2}-c^\dagger_{3j-1})|\bm{0}\rangle.
\end{equation*}

\subsubsection{Finite coupling}
Now we consider general values for $y$.
We shall normalise the polynomial coefficients in
the expansion $|\Psi(y)\rangle =\sum_ \alpha \psi_\alpha(y)|\alpha\rangle$ such
that the zero-order term for $ \alpha=010010010\cdots$ is $+1$.

As an example, we present the ground state for $N=6$ sites in the sector with $n=2$
particles. We find the amplitudes \begin{align*} 101000\quad & -y^5 & 100100\quad &
2 y^4 \left(1+y^2\right)\\ 010100\quad & -y^2 \left(1+2 y^2+2 y^4\right) &
100010\quad & -y^2 \left(1+y^2\right) \left(1+2 y^2\right)\\ 010010\quad & 1+4
y^2+4 y^4+2 y^6 & 001010\quad & -y \left(1+3 y^2+y^4\right)\\ 100001\quad & y^3
\left(1+2 y^2\right) & 010001\quad & -y \left(1+y^2\right) \left(1+2 y^2\right)\\
001001\quad & y^2 \left(1+3 y^2\right) & 000101\quad & -y^3 \end{align*}
We see that all components have integer coefficients and definite parity under $y\to -y $. The second property can be justified in a way which is similar to the one for staggering (I).
The projections on the asymptotic ground states are given by
\begin{align*}
\langle\Psi_P|\Psi(y)\rangle &=1+4 y^2+4 y^4+2 y^6,\\ \langle\Psi_S|\Psi(y)\rangle
&=1+6 y^2+11 y^4+8 y^6.
\end{align*} We find these two polynomials as factors in
the square norm. Indeed, for six sites
\begin{equation*}
  ||\Psi(y)||^2= \left(1+4
y^2+4 y^4+2 y^6\right) \left(1+6 y^2+11 y^4+8 y^6\right).
\end{equation*}
We have
verified these properties up to $n=8$ particles ($N=24$ sites), and identified the
projections on the asymptotic ground states in terms of polynomials associated with
the Hirota equation after suitable variable transformations. Thus, we formulate the
following

\begin{conjecture}
With the above normalisations all ground-state components are polynomials with integer coefficients and definite parity under $y\to -y$.
For all $n$ we have the projection
\begin{equation*}
\langle\Psi_P|\Psi(y)\rangle= 
\left(\frac{3+\zeta}{2}\right)^{n(n+1)} p_n\left(\frac{1-\zeta}{3+\zeta}\right) ,\quad \zeta^2=1+8y^2.
\end{equation*}
As $|\Psi_P\rangle$ is a single-component state, we can generate ground states for general $n$ with the help of the iterative algorithm and $m(y) =\langle\Psi_P|\Psi(y)\rangle$.
Likewise, the projection on the singlet state is given by
\begin{equation*}
  \langle \Psi_S|\Psi(y)\rangle = \left(\frac{\zeta}{2}\right)^{n(n+1)}q_n(\zeta^{-1})
  ,\quad \zeta^2=1+8y^2.
\end{equation*}
The square norm factorises into the product of these two projections
\begin{equation*}
  ||\Psi(y)||^2=\langle\Psi_P|\Psi(y)\rangle\langle \Psi_S|\Psi(y)\rangle.
\end{equation*}
\end{conjecture}

\subsubsection{Densities}
Let us again examine the
probability to find a particle at the first site. For $n=1,\dots,4$ we find that it
is scale-free: \begin{align*} y^{-4}\rho^{(1)}_1(y)&=\bm{1-3 y^2}+7 y^4-15 y^6+31
y^8-63 y^{10}+127 y^{12}+\dots \\ y^{-4}\rho^{(1)}_2(y)&=\bm{1-3 y^2+12
y^4-56 y^6}+246 y^8-1002 y^{10}+3884 y^{12}+\dots \\
y^{-4}\rho^{(1)}_3(y)&=\bm{1-3 y^2+12 y^4-56 y^6+288 y^8-1584 y^{10}}+8723
y^{12}+\dots\\
y^{-4}\rho^{(1)}_4(y)&=\bm{1-3 y^2+12 y^4-56 y^6+288y^8-1584 y^{10}+9152 y^{12}}+\dots
\end{align*}
We checked the
scale-free property up to $n=8$ particles. The stable coefficients are correctly
reproduced by the generating function
\begin{equation}
\rho^{(1)}_\infty(y)=\frac{1+12y^2+24y^4-(1+8y^2)^{3/2}}{32y^2},\quad y<1,
\label{eqn:bdyy1}
\end{equation}
which gives the correct value $\rho^{(1)}_\infty(y=1)=5/16$ at the
critical point. Around $y=\infty$ the quantity is scale free, too. We find
\begin{align*}
\rho^{(1)}_1(y)&= \bm{\frac{1}{2}}-\frac{3}{4 y^2}+\frac{7}{8
y^4}-\frac{15}{16 y^6}+ \frac{31}{32 y^8}+\dots\\
\rho^{(1)}_2(y)&=\bm{\frac{1}{2}-\frac{1}{8 y^2}}-\frac{65}{64 y^4}+\frac{1083}{512
y^6}-\frac{8601}{4096 y^8}+\dots \\ \rho^{(1)}_3(y)&=\bm{\frac{1}{2}-\frac{1}{8
y^2}-\frac{1}{32 y^4}}-\frac{27}{16 y^6}+ \frac{607}{128 y^8}+\dots\\
\rho^{(1)}_4(y)&=\bm{\frac{1}{2}-\frac{1}{8 y^2}-\frac{1}{32 y^4}-\frac{3}{256
y^6}}- \frac{12179}{4096 y^8}+\dots
\end{align*}
Again, we checked this up to $n=7$ particles, and inferred the infinite system limit:
\begin{equation*}
\rho_\infty^{(1)}(y)=\frac{1-4 y^2+8 y^4-8 y \left(y^2-1\right)^{3/2}}{16
y^2},\quad y \geq 1
\end{equation*}
Notice that this has the same structure as the
result for the case (I) staggering, and is compatible with the value $5/16$ at
$y=1$. This guess implies a discontinuity for the second and higher derivatives at
the critical point.

As similar pattern is found by probing if a particle is present on the last site.
The data is consistent with $\rho^{(N)}_n(y)=y^{-2}\rho_n^{(1)}(y)$ for finite $n\leq 8$.
We conjecture this to hold for arbitrary system sizes. The expectation values in the
thermodynamic limit are then readily deduced from our previous conjectures.

Let us now consider the centre of the chain. We focus on the density $\rho_n^{(j_c)}(y)$ with $j_c=3n/2$ for even $n$ and $j_c=3(n+1)/2$ for odd $n$. It turns out
that this quantity is scale-free both in the weak-coupling and the strong-coupling regime.
From the stable part of the corresponding series expansions we infer the expression for infinite system size:
\begin{equation*}
  \rho^{(j_c)}_\infty(y) =
  \begin{cases}
      \frac{1+2y^2-\sqrt{1+8y^2}}{2(y^2-1)}, & y\leq 1\\
    \frac{4y^2-1-4y\sqrt{y^2-1}}{1+8y^2}, & y\geq 1
  \end{cases}.
\end{equation*}
These coincide with the expectation values for the densities in the \textit{parity-odd} ground state for the staggered periodic chain.

\section{Ground state properties of the staggered chain with ${N=3n-1}$ sites}
\label{sec:3n1}

In this section we consider $N=3n-1$ sites. Despite the fact that the staggering is ``incommensurate'' we will point out that a relation to the Hirota polynomials continues to be present. Moreover, we shall show that the kink state appears in the weak-coupling limit of staggering (II) whereas the asymptotic ground states for staggering (I) have a very simple structure.

\subsection{Staggering I}
\subsubsection{Asymptotic ground states}
Let us focus on the chain with $N=3n-1$ sites and staggering $(y,1,y)$.
We determine the asymptotic ground states which are considerably
simpler than in the cases studied so far. For $y\to 0$ we seek for a state that is annihilated by $Q=\sum_{j=1}^n d_{3j-1}$ and its adjoint. So we know that for such a state every site $3j-1$ must be empty, but have a neighbouring site which is occupied. The site $3n-1$ has only one neighbour and therefore we must place a particle at site $3n-2$. Because of the hard-core exclusion there cannot be a particle on site $3n-3$. However, since the site $3(n-1)-1$ must have a particle in its neighbourhood we are left with the same problem as before, this time with the number of sites reduced by three, and the number of particles by one. We can thus proceed by iteration and see that the ground state is given up to a multiplicative factor by
\begin{equation*}
	|\Psi(y=0)\rangle = |\Psi_L\rangle = |100100\cdots10010\rangle.
\end{equation*}
Moreover, we see that the kink for $N=3n$ sites at $y=0$ basically appeared because we had the freedom to put a particle on the additional site $3n$.

For $y\to \infty$ the construction of the ground state is similarly straightforward. It must be annihilated by $Q= {d_1}+\sum_{j=1}^{n-1}d_{3j}+d_{3j-1}$ and its adjoint. As $Q$ acts on the isolated site $1$ we conclude that this one has to be empty but contain a particle in its neighbourhood, thus there must be a particle on site $2$. Because of the nearest-neighbour exclusion, there cannot be a particle on site $3$, nor is it possible to insert one there. Therefore the state is automatically annihilated by $d_3$ and its adjoint. We see that we are now left with the problem of placing $n-1$ particles on a chain with $3(n-1)-1$ according to the same rules as before. Again, we can proceed by iteration. Therefore, the only possible ground state at strong coupling is given up to normalisation by
\begin{equation*}
	|\Psi(y)\rangle\mathop{\sim}_{y\to \infty} |\Psi_R\rangle = |010010\cdots01001\rangle.
\end{equation*}
The simplicity of these asymptotic ground states will lead to simple scale-free expansions as we shall see. Moreover, notice that $|\Psi_L\rangle$ and $|\Psi_R$ represent exactly the two type of order which the kink state of section \ref{sec:asymy1y} interpolates.

\subsubsection{Finite coupling}
For $N=5$ sites, the ground state
components at finite coupling are given by
\begin{align*}
	10100& \quad -y^2& \quad
10010& \quad 1+2 y^2\\ 01010& \quad -y \left(1+2 y^2\right)& \quad 10001& \quad -y
\left(1+y^2\right)\\ 01001& \quad y^2 \left(1+2 y^2\right)& \quad 00101& \quad -y^3
\end{align*}
We see that all components are polynomials with integer coefficients which have a definite sign for a given configuration. The sign is flipped when going from one configuration to another which differs from the former by a single fermion move. The parity of the component polynomials under $y\to -y$ can be obtained by counting the number of particles on sites $3j-1$. This is related to a simple unitary transformation implementing this reflection, as explained previously.
The square norm is $||\Psi(y)||^2=\left(1+2 y^2\right) \left(1+4 y^2+4
y^4+2 y^6\right)$. We have $\langle \Psi_L|\Psi(y)\rangle = (1+2y^2)$ and
$\langle \Psi_R|\Psi(y)\rangle=y^2(1+2y^2)$, i.e. one of
the polynomials occurring in the square norm can be obtained as projections on
asymptotic ground states.

\begin{conjecture}
For general $n$ the ground-state components are polynomials with integer coefficients and definite parity under $y\to -y$. We have the projections
\begin{equation*}
  \langle \Psi_L|\Psi(y)\rangle = \left(\frac{\zeta}{2}\right)^{n(n-1)}q_{n-1}(\zeta^{-1}),\quad \zeta^2=1+8y^2
\end{equation*}
and
\begin{equation*}
  \langle \Psi_R|\Psi(y)\rangle = (-y)^n\langle \Psi_L|\Psi(y)\rangle.
\end{equation*}
As both $|\Psi_L\rangle$ and $|\Psi_R\rangle$ are single-component states, we can straightforwardly apply the iterative algorithm with $m(y) =\langle \Psi_L|\Psi(y)\rangle$ to generate general ground states with arbitrary $n$. These ground states are therefore characterised by a single polynomial.
The square norm factorises into two polynomials with integer coefficients:
\begin{equation*}
  ||\Psi(y)||^2= \langle \Psi_L|\Psi(y)\rangle\times \left(\frac{3+\zeta}{2}\right)^{n(n+1)} p_n\left(\frac{1-\zeta}{3+\zeta}\right) ,\quad \zeta^2=1+8y^2.
\end{equation*}
\end{conjecture}

\subsubsection{Densities}
The boundary densities at site $j=1$ have scale free series expansions both at $y=0$ and $y\to \infty$. From the stable parts of their series we conjecture their forms in the limit of infinite systems:
\begin{equation*}
	\rho_\infty^{(1)}(y)= 
	\begin{cases}
		\frac{1+44y^2-8y^4-(1+8y^2)^{3/2}}{32y^2}, & y\leq 1\\ \frac{1 + 12 y^2 - 8 y^4 + 8 y
(y^2 - 1)^{3/2}}{16 y^2}, & y\geq 1.
	\end{cases}
\end{equation*}
As anticipated earlier, the expression for $y\leq 1$ is equivalent to the one for $N=3n$ sites.

\subsection{Staggering II}
\label{sec:3n1yy1}
We now turn to staggering II with patter $(y,y,1)$. As it will turn out, for $N=3n-1$ this is the choice which leads to a kink state at $y=0$. The derivation of its precise form turns out to be slightly more complicated than the $N=3n$ case.

\subsubsection{Asymptotic ground states}
We start with $y=0$. The supercharge is given by $Q= Q_0=\sum_{j=1}^{n-1}d_{3j}$. We
focus on $n>1$. A state annihilated by $Q_0$ and $Q^ \dagger_0$ must not have
particles on sites $3j$ but at least one particle adjacent to them. We will call
this property ``adjacency condition''. We see immediately that it can only be fulfilled if there are at least $n-1$ particles, and at most $n$. The second sector
contains the ground states which will be compatible with non-zero values of $y$. The sites
$3j$ divide the system naturally into $n$ boxes of 2 sites. Here is an example for $n=3$:
\begin{equation*}
	{\boxed{\square\square}\,\bm{0}\,\boxed{\square\square}\,\bm{0}\,\boxed{\square\square}}
\end{equation*}
Given $n$ particles we need put a particle in each of them. Consider the leftmost of these boxes. If we put a particle on site
$1$ then there must be particles on all sites $3j-2$, $j=1,\dots,n$ in order to
satisfy the adjacency condition. However, if the particle in the leftmost box is at
site $2$, we already satisfy the adjacency condition for site $3$. For the
remaining $n-1$ boxes we are left with the same problem as before, but with the
number of sites reduced by three. This reminds us of the kink state problem
which we encountered for the staggering $(y,1,y)$ in section \ref{sec:asymy1y}: in fact, any of the following $n+1$ \textit{kink states}
\begin{subequations}
\begin{align}
	|m\rangle
&=|\underset{m}{\underbrace{01\bm{0}01\bm{0}\cdots 01\bm{0}}} \underset{n-1-m}
{\underbrace{10\bm{0} 10\bm{0} \cdots 10\bm{0}}} 10\rangle, \quad m=0,\dots,n-1,\\
|n\rangle &=|01\bm{0}010\bm{0}\cdots 01\bm{0}01\rangle
\end{align}
\label{eqn:kinkstates}
\end{subequations}
is annihilated by the supercharges.
We see that for $1\leq m \leq n-1$ the state $|m\rangle$ contains a defect of type $\cdots 01\bm{0}10\bm{0}\cdots$. We expect to ground state of our problem at $y=0$ to be a linear combination:
\begin{equation*}
	|\Psi_0\rangle = \sum_{m=0}^{n}a_m|m\rangle,
\end{equation*}
where the $a_m$ are complex numbers to be determined from continuity to arbitrary values of the coupling. For finite $y$ the wave function will be a
polynomial $|\Psi(y)\rangle=|\Psi_0\rangle+y|\Psi_1\rangle+\dots$ with relations
$Q_0| \Psi_1\rangle+Q_1|\Psi_0\rangle=0$, and
$Q_0^\dagger|\Psi_1\rangle+Q_1^\dagger|\Psi_0\rangle=0$ where
$Q_1=\sum_{j=1}^n(d_{3j-2}+d_{3j-1})$. All states in $|\Psi_0\rangle$ are already
annihilated by $Q_1^\dagger$. So we concentrate on the action of $Q_1$ on
$|\Psi_0\rangle$. This operator takes out particles from the $n$
boxes introduced previously. Thus, it is useful to introduce states
$|m,j\rangle,\,j=1,\dots,n$ which correspond to the state $|m\rangle$ with the
particle from the $j$-th box removed. This definition is somewhat redundant
because of the identity $|m-1,m\rangle = |m,m\rangle$ for $1\leq m\leq n$. So let us write
explicitly
\begin{align*}
	Q_1|\Psi_0\rangle=\sum_{j=1}^n\sum_{m=0}^n (-1)^{j-1}a_m|m,j\rangle
\end{align*}
where the sign comes from the fermionic character of the particles.
As we wish to solve the equation
$Q_0|\Psi_1\rangle+Q_1|\Psi_0\rangle=0$ we should ask if a state $|m,j\rangle$ can
be created by acting with $Q_0$ on something (i.e. if it is $Q_0$-exact). First, for $m=0$ this is possible if
$j=2,3,\dots,n$ because only then there will be an empty site $3(j-1)$ with no particle
on a neighbouring site. The illustration
\begin{align*}
	|0,1\rangle &=
|00\bm{0}10\bm{0}10\bm{0}\cdots 10\bm{0}10\rangle\\ |0,2\rangle &=
|10\underset{\uparrow}{\bm{0}}00\bm{0}10\bm{0}\cdots 10\bm{0}10\rangle\\
|0,3\rangle &= |10\bm{0}10\underset{\uparrow}{\bm{0}}00\bm{0}10\bm{0}\cdots
10\bm{0}10\rangle\\ \vdots
\end{align*}
points out that the state $|0,1\rangle$ cannot be created through the action of $Q_0$.
A very similar argument holds for
$|n,n\rangle$ which cannot be obtained through $Q_0$ acting on some other state,
whereas the $|n,j\rangle,\,1<j<n$ can. Let us now turn to the more general $1\leq m \leq n-1$. A
state $|m,j\rangle$ can be created through the action of $Q_0$ if $j\neq m,m+1$.
Again this is because in these cases removing particles will create an empty site
$3j$ with no particle on a neighbouring site. If however $j=m$ or $j=m+1$ then 
there must be a particle on one of the sites adjacent to $3j$, and so in
this case a creation through action of $Q_0$ is not possible.

Let us now write $Q_1|\Psi_0\rangle = |\Psi_0'\rangle +|\Psi_0''\rangle$ where
$|\Psi_0'\rangle$ collects all states which can be created through the action of
$Q_0$ on something, and $|\Psi_0''\rangle$ those which cannot (in mathematical parlance, $|\Psi_0'\rangle$ is $Q_0$-exact, while $|\Psi_0''\rangle$ is not). By definition, we
must have $|\Psi_0''\rangle=0$. More explicitly, we find
\begin{align*}
|\Psi_0''\rangle=0=&\sum_{m=1}^{n-1}\left((-1)^{m-1}a_m|m,m\rangle+(-1)^{m}a_m|m,m+1\rangle\right)\\
&+a_0|0,1\rangle+(-1)^{n-1}a_n|n,n\rangle.
\end{align*}
Using now the rule $|m-1,m\rangle = |m,m\rangle$, this can be reduced to \begin{equation*}
	\sum_{m=1}^n (-1)^{m-1}(a_{m-1}+a_m)|m,m\rangle=0.
\end{equation*}
Because all states in this sum are linearly independent, we find that
$a_0=-a_1=a_2=-a_3=\dots$. We fix the normalisation by setting $a_0=1$. This
completes the construction of the ground state at $y=0$ and shows that it is again
a superposition of kink states, in this case however with alternating coefficients. Hence, we found\begin{equation}
	|\Psi(y=0)\rangle=|\Psi_K\rangle = \sum_{m=0}^{n}(-1)^m |m\rangle.
	\label{eqn:altkink}
\end{equation}

The ground state in the limit $y\to \infty$ is simpler. It is
annihilated by $Q_1$ and $Q_1^\dagger$, which act on sites $3j-2$ and $3j-1$ with
$j=1,\dots,n$, and thus on the $n$ boxes introduced previously. We
know that by continuity we must look in the sector with $n$ particles. As the boxes are
separate, we can try to solve the problem
$(d_{3j-2}+d_{3j-1})|\chi\rangle=(d_{3j-2}^\dagger
+d_{3j-1}^\dagger)|\chi\rangle=0$ first and then take appropriate tensor products.
From section \ref{sec:asymyy1} we already know that this problem is solved by a singlet state $\cdots 10
\cdots - \cdots 01 \cdots$. Therefore we conclude that the $y\to\infty$ ground state is
given by
\begin{equation*}
	|\Psi(y)\rangle \mathop{\sim}_{y\to \infty}|\Psi_S\rangle = \prod_{j=1}^{n-1}(c_{3j-2}^\dagger -c_{3j-1}^\dagger)|\bm{0}\rangle.
\end{equation*}

\subsubsection{Finite coupling}
Let us finally turn to the case of finite coupling. For $N=5$ sites, the ground
state components are given by
\begin{align*}
	10100& \quad -y & \quad 10010& \quad
1+2 y^2\\ 01010& \quad -(1+2 y^2) & \quad 10001& \quad -2 y^2\\ 01001& \quad 1+2
y^2 & \quad 00101& \quad -y
\end{align*}
Together with the square norm
$||\Psi(y)||^2=(1+2y^2)(3+8y^2)$. We have $\langle \Psi_K|\Psi(y)\rangle =
3(1+2y^2)$ and $\langle\Psi_S|\Psi(y)\rangle=3+8y^2 $, i.e. both polynomials
occurring in the square norm can be obtained as projections on asymptotic ground
states. Moreover, we see that the state given here is even under the parity operation. The fact, that we have definite parity is clear because for the present staggering and number of sites the Hamiltonian commutes with parity $[H,P]=0$. The parity of the ground state at finite $y$ can thus be read off from the asymptotic cases discussed in the last section. Upon inspection, we find thus that the ground state for this staggering at $N=3n-1$ sites has parity $(-1)^n$ (notice that as opposed to the case of periodic chains studied in \cite{fendley:10_1}, the singlet state at $y\to \infty$ can be even under parity, namely if $n$ is even).

\begin{conjecture}
For general $n$ all ground-state components are polynomials with integer coefficients and definite parity under $y\to -y$. The projection on the kink state \eqref{eqn:altkink} is given by
\begin{equation}
  \langle \Psi_K|\Psi(y)\rangle = (n+1)m(y),\quad m(y)= \left(\frac{\zeta}{2}\right)^{n(n-1)}q_{n-1}(\zeta^{-1}),\quad \zeta^2={1+8y^{2}}.
\end{equation}
The structure of the wave function is given by
\be
|\Psi(y)\rangle = m(y) \, |\Psi_{K}\rangle + |\widetilde \Psi\rangle, \qquad \langle\Psi_{K}|\widetilde\Psi\rangle=0,
\nonumber
\ee
and thus considerably simpler than for the kink state of staggering (I), what allows for a straightforward application of the iterative algorithm. The square norm factorises into two polynomials with positive integer coefficients which coincide with the projections on the kink and singlet state respectively.
\begin{equation*} 
    ||\Psi(y)||^2 = \frac{1}{n+1}\langle\Psi(y)|\Psi_K\rangle \langle \Psi_S|\Psi(y)\rangle
\end{equation*}
The projection on the singlet state $\langle \Psi_S|\Psi(y)\rangle$ yields the unknown polynomials $t_n(y)$ introduced in section \ref{sec:3n}.
\end{conjecture}

\subsubsection{Densities}
We investigate again the boundary density $\rho_n^{(1)}(y)$. As the physics of weak-coupling regime is governed by the kink state $|\Psi_K\rangle$ we expect the density not to be scale-free at $y=0$. Indeed, up to $n=8$ particles the series expansion is rather compatible with the series
\begin{align*}
  \rho_n^{(1)}(y) =& \frac{1}{n+1}+\frac{n-1}{(n+1)^2}y^2+\frac{n(n-1)(n+3)}{(n+1)^3}y^4-\frac{3n^4+9n^3+15n^2+3n+2}{(n+1)^4}y^6\nonumber\\
  &+\frac{4(3n^5+12n^4+18n^3+20n^2+11n+4)y^8}{(n+1)^5}+\dots
\end{align*}
In the infinite-system limit $n\to \infty$ we thus find
\begin{equation*}
  \rho_\infty^{(1)}(y) = y^4-3y^6+12y^8+\dots
\end{equation*}
what matches the first terms of the series expansion at $y=0$ of the corresponding results for $N=3n$ sites given in \eqref{eqn:bdyy1}. Based on physical grounds, we conjecture that our the boundary density considered here converges to the expressions given there.

Conversely, the boundary density in the $y>1$ regime displays at scale-free behaviour around the point $y\to \infty$. From the stable part of the series expansion we conjecture the expression
\begin{equation*}
  \rho^{(1)}_\infty(y)=\frac{1-4 y^2+8 y^4-8 y (y^2-1)^{3/2}}{16 y^2}.
\end{equation*}

\section{Conclusion}
\label{sec:concl}

In conclusion, we presented several conjectures on the ground-state properties of a staggered one-dimensional supersymmetric fermion model with open boundary conditions, based on the exact diagonalisation of small systems. In particular, we conjecture a relation between certain ground state components of the model, and a class of polynomials which appeared in the literature on the XYZ spin chain along a special line of couplings. The very same polynomials occur in the square norm of the corresponding ground state vectors. Moreover, exploiting the scale-free property we determined exact expressions for local site-occupation probabilities for infinite systems.

Given these observations it is quite natural to ask if there are open boundary conditions for the XYZ spin chain, possibly amongst the integrable ones \cite{inami:94}, which still lead to exact zero-energy ground states with similar properties to the ones studied here. This is the case for the XXZ chain as was shown in \cite{yang:04}, yet it might be more difficult off criticality because one can show that the underlying supersymmetry in the spin chains does not survive this deformation \cite{hagendorf:12_1}. Another interesting open problem concerns the choice of the coupling constants $\lambda_j$: the results presented here relate to a staggered pattern. In fact, it turns out that some of the conjectured sum rules survive for generic choices. In fact, one may show that these follow from the supersymmetry alone. This problem will be addressed in \cite{hagendorf:2bp}.

\subsection*{Acknowledgments}
This work was supported by the ERC AG CONFRA and by the Swiss NSF, and in part by the NSF grants DMR/MSPA-0704666 and PHY05-51164. CH would like to thank the Kavli Institute for Theoretical Physics, were this work was started, for hospitality. We thank Paul Fendley for many very stimulating discussions. CH thanks Liza Huijse for a very fruitful correspondence and many discussions on the subject.

\appendix

\section{Polynomials}
In this appendix, we list various polynomials encountered in the text $p_k(w)$ and $q_k(w)$. They can also be found in \cite{mangazeev:10}.
The first explicit cases of the $p_{k}(w)$ polynomials are 
\begin{eqnarray*}
p_{-4}(w) &=& 1-4 w+22 w^2-70 w^3+245 w^4-680 w^5+1940 w^6-3260 w^7+5915 w^8 \nonumber \\
&& -6500 w^9+7878 w^{10}-3822 w^{11}+2431 w^{12}, \nonumber \\ 
p_{-3}(w) &=& 1-3 w+12 w^2-30 w^3+81 w^4-63 w^5+66 w^6, \nonumber \\ 
p_{-2}(w) &=& 1-2 w+5 w^2, \nonumber \\ 
p_{-1}(w) &=& 1, \nonumber \\ 
p_{0}(w) &=& 1, \\ 
p_{1}(w) &=& 1+w+2 w^2, \nonumber \\ 
p_{2}(w) &=& 1+2 w+7 w^2+10 w^3+21 w^4+12 w^5+11 w^6, \nonumber \\ 
p_{3}(w) &=& 1+3 w+15 w^2+35 w^3+105 w^4+195 w^5+435 w^6+555 w^7+840 w^8 \nonumber\\
&& +710 w^9+738 w^{10}+294 w^{11}+170 w^{12}, \nonumber \\ 
p_{4}(w) &=& 1+4 w+26 w^2+82 w^3+319 w^4+840 w^5+2488 w^6+5572 w^7+13524 w^8\nonumber \\
&& +24920 w^9+48776 w^{10}+72800 w^{11}+114716 w^{12}+135464 w^{13}+169536 w^{14} \nonumber \\
&& +148972 w^{15}+141835 w^{16}+85044 w^{17}+58406 w^{18}+17822 w^{19}+7429 w^{20}. \nonumber
\end{eqnarray*}
The first explicit cases of the $q_{k}(w)$ polynomials are 
\begin{eqnarray*}
q_{-4}(w) &=& 1+8 w^2+35 w^4+560 w^6+1215 w^8+1848 w^{10}+429 w^{12}, \nonumber \\ 
q_{-3}(w) &=& 1+3 w^2+39 w^4+21 w^6, \nonumber \\ 
q_{-2}(w) &=& 1+3 w^2, \nonumber \\ 
q_{-1}(w) &=& 1, \nonumber \\ 
q_{0}(w) &=& 1, \\ 
q_{1}(w) &=& 1+3 w^2, \nonumber \\ 
q_{2}(w) &=& 1+8 w^2+29 w^4+26 w^6, \nonumber \\ 
q_{3}(w) &=& 1+15 w^2+112 w^4+518 w^6+1257 w^8+1547 w^{10}+646 w^{12}, \nonumber \\ 
q_{4}(w) &=& 1+24 w^2+291 w^4+2338 w^6+13524 w^8+54474 w^{10}+150472 w^{12}+276678 w^{14}
\nonumber \\
&& +312195 w^{16}+192694 w^{18}+45885 w^{20}. \nonumber  
\end{eqnarray*}


\begin{thebibliography}{10}
\bibitem{fendley:03_2}
P.~{Fendley}, K.~{Schoutens}  and J.~{de Boer},
\newblock {\em {Lattice Models with N=2 Supersymmetry}},
\newblock Phys. Rev. Lett. {\textbf{90}} (2003)   120402,
\newblock arXiv:hep-th/0210161.

\bibitem{fendley:03}
P.~{Fendley}, B.~{Nienhuis}  and K.~{Schoutens},
\newblock {\em {Lattice fermion models with supersymmetry}},
\newblock J. Phys. A: Math. Gen. {\textbf{36}} (2003)   12399--12424,
\newblock arXiv:cond-mat/0307338.

\bibitem{beccaria:05}
M.~{Beccaria} and G.~F. {de Angelis},
\newblock {\em {Exact Ground State and Finite-Size Scaling in a Supersymmetric
  Lattice Model}},
\newblock Phys. Rev. Lett. {\textbf{94}} {\textbf{10}} (2005)   100401,
\newblock arXiv:cond-mat/0407752.

\bibitem{razumov:00}
A.~V. {Razumov} and Y.~G. {Stroganov},
\newblock {\em {Spin chains and combinatorics}},
\newblock J. Phys. A : Math. Gen. {\textbf{34}} (2001)   3185--3190,
\newblock arXiv:cond-mat/0012141.

\bibitem{razumov:01}
A.~V. {Razumov} and Y.~G. {Stroganov},
\newblock {\em {Spin chains and combinatorics: twisted boundary conditions}},
\newblock J. Phys. A: Math. Gen. {\textbf{34}} (2001)   5335--5340,
\newblock arXiv:cond-mat/0102247.

\bibitem{degier:02}
J.~{de Gier}, M.~T. {Batchelor}, B.~{Nienhuis}  and S.~{Mitra},
\newblock {\em {The XXZ spin chain at {$\Delta=-1/2$}: Bethe roots, symmetric
  functions, and determinants}},
\newblock J. Math. Phys. {\textbf{43}} (2002)   4135--4146,
\newblock arXiv:math-ph/0110011.

\bibitem{difrancesco:06}
P.~{Di Francesco}, P.~{Zinn-Justin}  and {J.-B.} {Zuber},
\newblock {\em {Sum rules for the ground states of the O(1) loop model on a
  cylinder and the XXZ spin chain}},
\newblock J. Stat. Mech. {\textbf{8}} (2006)  ~11,
\newblock arXiv:math-ph/0603009.

\bibitem{yang:04}
X.~{Yang} and P.~{Fendley},
\newblock {\em {Non-local spacetime supersymmetry on the lattice}},
\newblock J. Phys. A: Math. Gen. {\textbf{37}} (2004)   8937--8948,
\newblock arXiv:cond-mat/0404682.

\bibitem{fendley:05}
P.~{Fendley}, K.~{Schoutens}  and H.~{van Eerten},
\newblock {\em {Hard squares with negative activity}},
\newblock J. Phys. A: Math. Gen. {\textbf{38}} (2005)
  315--322,
\newblock arXiv:cond-mat/0408497.

\bibitem{huijse:08_2}
L.~{Huijse}, J.~{Halverson}, P.~{Fendley}  and K.~{Schoutens},
\newblock {\em {Charge Frustration and Quantum Criticality for Strongly
  Correlated Fermions}},
\newblock Phys. Rev. Lett. {\textbf{101}} (2008)   146406,
\newblock arXiv:0804.0174.

\bibitem{fendley:05_2}
P.~{Fendley} and K.~{Schoutens},
\newblock {\em {Exact Results for Strongly Correlated Fermions in 2+1
  Dimensions}},
\newblock Phys. Rev. Lett. {\textbf{95}} {\textbf{4}} (July 2005)
  046403,
\newblock arXiv:cond-mat/0504595.

\bibitem{jonsson:06}
J.~Jonsson,
\newblock {\em {Hard Squares with Negative Activity and Rhombus Tilings of the
  Plane}},
\newblock Electr. J. Comb. {\textbf{13}} (2006) R67.

\bibitem{jonsson:10}
J.~Jonsson,
\newblock {\em Certain homology cycles of the independence complex of grids},
\newblock Discrete Comput. Geom. {\textbf{43}} (2010)   927--950.

\bibitem{huijse:10_1}
L.~Huijse and K.~Schoutens,
\newblock {\em {Supersymmetry, lattice fermions, independence complexes and
  cohomology theory}},
\newblock Adv. Theor. Math. Phys. {\textbf{14}} (2010)   643--694.

\bibitem{huijse:10}
L.~Huijse,
\newblock {\em A supersymmetric model for lattice fermions},
\newblock PhD thesis, Universiteit van Amsterdam (2010),
\newblock available at \href{http://dare.uva.nl/de/record/341605}{http://dare.uva.nl/de/record/341605}

\bibitem{huijse:11_2}
L.~{Huijse},
\newblock {\em {Detailed analysis of the continuum limit of a supersymmetric
  lattice model in 1D}},
\newblock J. Stat. Mech. (2011)   P04004,
\newblock arXiv:1102.1700.

\bibitem{fendley:10}
P.~{Fendley} and C.~{Hagendorf},
\newblock {\em {Exact and simple results for the XYZ and strongly interacting
  fermion chains}},
\newblock J. Phys. A: Math. Theor. {\textbf{43}} (2010)   402004.

\bibitem{fendley:10_1}
Paul Fendley and Christian Hagendorf,
\newblock {\em Ground-state properties of a supersymmetric fermion chain},
\newblock J.Stat.Mech. {\textbf{1102}} (2011)   P02014.

\bibitem{bazhanov:06}
V.~V. {Bazhanov} and V.~V. {Mangazeev},
\newblock {\em {The eight-vertex model and Painlev{\'e} VI}},
\newblock J. Phys. A: Math. Gen. {\textbf{39}} (2006)   12235--12243,
\newblock arXiv:hep-th/0602122.

\bibitem{mangazeev:10}
V.~V. {Mangazeev} and V.~V. {Bazhanov},
\newblock {\em {The eight-vertex model and Painlev{\'e} VI equation II:
  eigenvector results}},
\newblock J. Phys. A: Math. Theor. {\textbf{43}} (2010)   085206,
\newblock arXiv:0912.2163.

\bibitem{zinnjustin:12}
P.~{Zinn-Justin},
\newblock {Sum rule for the eight-vertex model on its combinatorial line},
\newblock arXiv:1202.4420 2012.

\bibitem{hagendorf:12}
Christian Hagendorf and Paul Fendley,
\newblock {\em {The eight-vertex model and lattice supersymmetry}},
\newblock J. Stat. Phys. {\textbf{146}} (2012)   1122--1155.

\bibitem{huijse:11}
L.~{Huijse}, N.~{Moran}, J.~{Vala}  and K.~{Schoutens},
\newblock {\em {Exact ground states of a staggered supersymmetric model for
  lattice fermions}},
\newblock Phys. Rev. B {\textbf{84}} (2011)   115124,
\newblock arXiv:1103.1368.

\bibitem{witten:82}
E.~Witten,
\newblock {\em {Constraints on supersymmetry breaking}},
\newblock Nucl. Phys. B {\textbf{202}} (1982)   253 -- 316.

\bibitem{hagendorf:12_1}
C.~Hagendorf,
\newblock {\em {Spin chains with dynamical lattice supersymmetry}},
\newblock arXiv:1207.0357 (2012).

\bibitem{okamoto:87}
K.~Okamoto,
\newblock {\em {Studies on the Painlev{\'e} equations. I: Sixth Painlev{\'e}
  equation PVI}},
\newblock Annali di Matematica pura ed applicata {\textbf{146}} (1987)
  337--381.

\bibitem{inami:94}
T.~Inami and H.~Konno,
\newblock {\em {Integrable XYZ spin chain with boundaries}},
\newblock J. Phys. A : Math. Gen. {\textbf{27}} (1994)   L913--L918.

\bibitem{hagendorf:2bp}
C.~Hagendorf,
\newblock in preparation (2012).
\end{thebibliography}
\end{document}